\documentclass{article}
\usepackage{spconf,graphicx}
\usepackage{amsfonts}
\usepackage{amsmath,amssymb,amsthm}
\usepackage{color}
\usepackage{makecell}
\usepackage{arydshln}
\usepackage{multirow}
\usepackage{multicol}

\def\0{{\mathbf 0}}
\def\1{{\mathbf 1}}
\def\a{{\mathbf a}}

\def\d{{\mathbf d}}

\def\f{{\mathbf f}}
\def\g{{\mathbf g}}
\def\l{{\mathbf l}}

\def\n{{\mathbf n}}

\def\x{{\mathbf x}}

\def\y{{\mathbf y}}
\def\z{{\mathbf z}}

\def\H{{\mathbf H}}

\def\L{{\mathbf L}}
\def\M{{\mathbf M}}
\def\P{{\mathbf P}}

\def\W{{\mathbf W}}
\def\X{{\mathbf X}}
\def\Y{{\mathbf Y}}

\def\cR{{\mathcal R}}

\def\rPr{{\mathrm{Pr}}}

\def\ie{{\textit{i.e.}}}

\def\rPr{{\textrm{Pr}}}

\title{3D Point Cloud Enhancement using Graph-Modelled Multiview Depth Measurements}
\name{Xue Zhang$^{\star}$, Gene Cheung$^{\star}$, Jiahao Pang$^{\dagger}$, Dong Tian$^{\dagger}$}
\address{$^{\star}$Dept of EECS, York University, Toronto, Canada \\ 
$^{\dagger}$ InterDigital, Princeton, New Jersey, USA}
\begin{document}
\ninept
\maketitle
\begin{abstract}
A 3D point cloud is often synthesized from depth measurements collected by sensors at different viewpoints. 
The acquired measurements are typically both coarse in precision and corrupted by noise. 
To improve quality, previous works denoise a synthesized 3D point cloud \textit{a posteriori} after projecting the imperfect depth data onto 3D space.
Instead, we enhance depth measurements on the sensed images \textit{a priori}, exploiting inherent 3D geometric correlation across views, before synthesizing a 3D point cloud from the improved measurements.
By enhancing closer to the actual sensing process, we benefit from optimization targeting specifically the depth image formation model, before subsequent processing steps that can further obscure measurement errors.
Mathematically, for each pixel row in a pair of rectified viewpoint depth images, we first construct a graph reflecting inter-pixel similarities via metric learning using data in previous enhanced rows. 
To optimize left and right viewpoint images simultaneously, we write a non-linear mapping function from left pixel row to the right based on 3D geometry relations.
We formulate a MAP optimization problem, which, after suitable linear approximations, results in an unconstrained convex and differentiable objective, solvable using fast gradient method (FGM).
Experimental results show that our method noticeably outperforms recent denoising algorithms that enhance after 3D point clouds are synthesized.
\end{abstract}
\section{Introduction}
\label{sec:intro}
\textit{Point Cloud} (PC) is a signal representation composed of discrete geometric samples of a physical object in 3D space, useful for a range of imaging applications such as immersive communication and virtual / augmented reality (AR/VR) \cite{wien2019standardization, steven2016virtual}. 
With recent advance and ubiquity of inexpensive active sensing technologies like Microsft Kinect and Intel RealSense, one method to generate a PC is to deploy multiple depth sensors to capture depth measurements (in the form of images) of an object from different viewpoints, then project these measurements to 3D space to synthesize a PC \cite{hartley2003multiple, huang2018apolloscape}.
Limitations in the depth acquisition process mean that the acquired depth measurements suffer from both imprecision and additive noise.
This results in a noisy synthesized PC, and previous works focus on denoising PCs using a variety of methods: low-rank prior, low-dimensional manifold model (LDMM), surface smoothness priors expressed as graph total variation (GTV), graph Laplacian regularizer (GLR), feature graph Laplacian regularizer (GFLR), etc \cite{dinesh2018local, zeng20193d, hu19}. 

However, all the aforementioned denoising methods enhance a PC \textit{a posteriori}, \ie, after a PC is synthesized from corrupted depth measurements. 
Recent work in image denoising \cite{punnappurath2019learning, nguyen2016raw} has shown that by denoising raw sensed RGB measurements directly on the Bayer-patterned grid \textit{before} demosaicking, contrast boosting and other steps typical in an image construction pipeline \cite{farsiu2005multiframe, sahu2017contrast} that obscure acquisition noise, one can dramatically improve the denoising performance compared to denoising on the image constructed \textit{after} the pipeline (up to 15dB in PSNR).
Inspired by these work, we propose to enhance\footnote{We call our processing an ``enhancement" that performs joint denoising and dequantization based on our depth image formation model.} measurements in acquired depth images \textit{a priori}, before projection to synthesize a PC.
In our case, by enhancing closer to the actual physical sensing process before various steps in a PC synthesis pipeline including registration, stitching and filtering, we benefit from optimization that directly targets our depth-sensor-specific image formation model with a finite-bit pixel representation.

Specifically, towards a graph-smoothness signal prior \cite{ortega18ieee,cheung18,pang17}, for each pixel row in a pair of rectified viewpoint depth images, we first construct a sparse graph reflecting inter-pixel similarities via metric learning \cite{hu19} using data in previous enhanced rows. 
To exploit inter-view correlation and optimize left and right viewpoint images simultaneously, we write a non-linear mapping function from the left pixel row to the right based on 3D geometry relations.
Using a depth image formation model that accounts for both additive noise and quantization, we formulate a \textit{maximum a posteriori} (MAP) optimization problem, which, after suitable linear approximations, results in an unconstrained convex and differentiable objective, solvable using fast gradient method (FGM) \cite{nesterov2013introductory}.
Experimental results show that by enhancing measurements at the depth image level, our method outperforms several recent PC denoising algorithms \cite{guennebaud2007algebraic, oztireli2009feature, mattei2017point} in two commonly used PC error metrics \cite{tian2017geometric}.

\vspace{0.02in}
\noindent
\textbf{Related Work}: 
Previous work on depth image enhancement \cite{hu16spl, gu2017learning, jeon2018reconstruction} typically enhances one depth map at a time using image-based signal priors.
When given two (or more) viewpoint depth maps, by ignoring the inherent cross-correlation between the views and optimizing each separately, the resulting quality is sub-optimal.
One exception is \cite{wan15}, which considers noiseless but quantized observations per pixel from two views as signals in specified quantization bins.
To reconstruct, the most likely signal inside both sets of quantization bins is chosen. 
Our work differs from \cite{wan15} in that 
our image formation model considers both additive noise and quantization, leading to a more challenging MAP problem involving likelihood and prior terms from both views.
We address this using appropriate linear approximations and FGM.



\section{System Overview}
\label{sec:system}
\begin{figure}
\begin{center}
\includegraphics[width=3in]{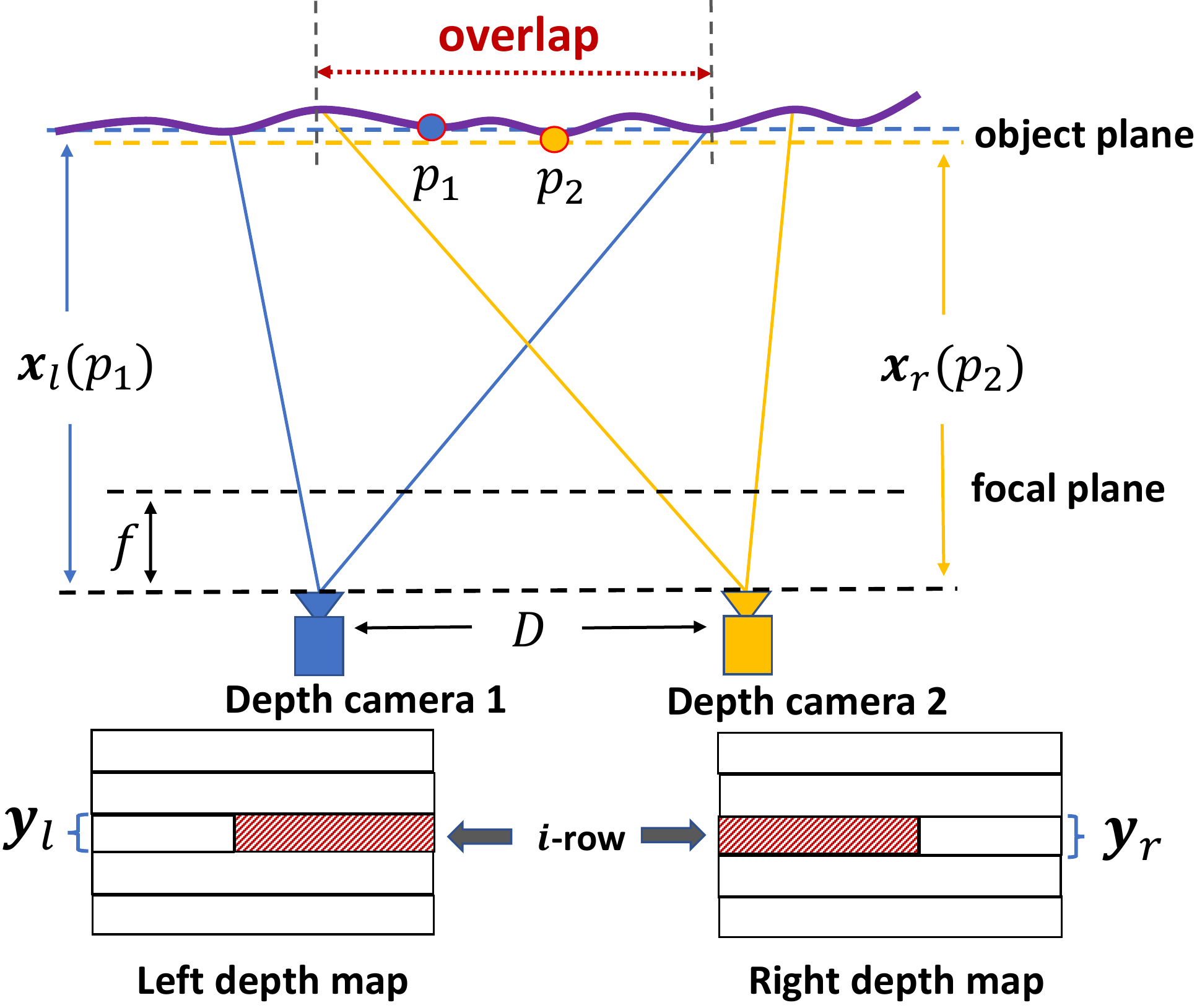}
\vspace{-0.1in}
\caption{\footnotesize An example of the camera system.}
\label{fig:model}
\end{center}
\vspace{-0.5cm}
\end{figure}

We assume a capturing system where the same 3D object is observed by two consumer-level depth cameras from different viewpoints, separated by distance $D$. 
Specifically, there exist overlapping fields of view (FoV) from the two cameras, so that there are multiple observations of the same 2D object surface.  
See Fig.\,\ref{fig:model} for an illustration. 
Each depth camera returns as output a depth map of resolution $H \times W$ and finite precision: each pixel is a noise-corrupted observation of the physical distance between the camera and the object, quantized to a $B$-bit representation. Without access to the underlying hardware pipeline, we assume that the depth map is the ``rawest" signal we can acquire from the sensor.

For simplicity, we assume that the two captured depth maps are \textit{rectified}; \ie, pixels in a row $i$ in the left view are capturing the same horizontal slice of the object as pixels in the same row $i$ in the right view. 
Rectification is a well-studied computer vision problem, and a known procedure \cite{loop1999computing} can be executed as a pre-processing step prior to our enhancement algorithm. 

\section{Problem Formulation}
\label{sec:formulate}
We first describe a depth-sensor-specific image formation model and a mapping from left-view pixels to right-view pixels.
We next define likelihood term and signal prior for depth images.
Finally, we formulate a MAP optimization problem to enhance multiview depth measurements.

\subsection{Image Formation Model}

Denote by $\y_l \in \mathbb{R}^N$ ($\y_r \in \mathbb{R}^N$) an observed depth pixel row $i$ in the left (right) view. 
See Fig.\,\ref{fig:model} for details.
(We forego index $i$ in $\y_l$ in the sequel for notation simplicity.)
Observed pixels are noise-corrupted and quantized versions of the original depth measurements, $\x_l^o$ and $\x_r^o$, respectively. 
Specifically, observation $\y_l$ and true signal $\x_l^o$ are related via the following formation model\footnote{Practical quantization of depth measurements into $B$-bit representation for existing sensors are often non-uniform: larger depth distances are quantized into larger quantization bins. For simplicity, we model uniform quantization here, but the non-uniform generalization is straightforward.}:
\begin{equation}
\y_l=\text{round} \left( \frac{\x_l^o+\n_l}{Q} \right) Q
\end{equation}
where $Q \in \mathbb{R}^+$ is a quantization parameter, and $\n_l \in \mathbb{R}^N$ is a zero-mean additive noise. 
The same formation model applies for the right view.
The goal is to optimally reconstruct signal $\widetilde{\X}=\{\widetilde{\x}_l, \widetilde{\x}_r\}$ given observation $\Y=\{\y_l, \y_r\}$. 

\subsection{View-to-view Mapping}

Pixel rows $i$ of the rectified left and right views, $\x_l$ and $\x_r$, are projections from the same 2D object surface onto two different camera planes, and thus are related. 
For simplicity, we assume that there is no occlusion when projecting a 3D object to the two camera views.
We employ a known 1D warping procedure \cite{jin2016region} to relate $\x_l$ and $\x_r$.  
For the $j$-th pixel in the left view, $x_{l,j}$, its (non-integer) horizontal position $s(j,x_{l,j})$ in the right view after projection is
\begin{align}
&s(j, x_{l,j}) = j-\delta(x_{l,j}) \nonumber \\
& \delta(x_{l,j}) = \frac{ f D}{x_{l,j}} 
\label{eq:disp}
\end{align}
where $\delta(\,)$ is the disparity, and $f$ is the camera focal length. 
Note that $s$ is a function of both left pixel's integer horizontal position $j$ and depth value $x_{l,j}$.

Assuming that the object surface is smooth, we interpolate right pixel row $\x_r$ given left pixel row $\x_l$ as
\begin{align}
\x_r=\W(\x_l) \, \x_l= \g(\x_l)
\label{eq:interpolate}
\end{align}
where $\W(\x_l) \in \mathbb{R}^{N \times N}$ is a real weight matrix. 
\eqref{eq:interpolate} states that $\x_r$ is linearly interpolated from $\x_l$ using weights $\W(\x_l)$, where $\W(\x_l)$ is a function of signal $\x_l$.
In particular, we model the weight $\omega_{ij}$ between right pixel $x_{r,i}$ and left pixel $x_{l,j}$ as

\vspace{-0.1in}
\begin{small}
\begin{align}
\omega_{ij}=&\frac{1}{\sum_{m=1}^{N} \exp \left(- \frac{(s(m,x_{l,m})-i)^2}{\sigma_s^2}\right)} \exp \left(- \frac{(s(j,x_{l,j})-i)^2}{\sigma_s^2}\right) 
 \label{eq:weight}
\end{align}
\end{small}
In words, weight $\omega_{ij}$ is larger if the distance between the projected position $s(j,x_{l,j})$ of left pixel $j$ and the target pixel position $i$ in $\x_{r}$ is small.
To simplify \eqref{eq:weight}, we assume a constant $C_i=1/\sum_{m=1}^{N} \exp \left(-\frac{(s(m, x_{l,m})-i)^2}{\sigma_s^2}\right)$. 
Combining with \eqref{eq:disp}, \eqref{eq:weight} is rewritten as
\begin{align}
 \omega_{ij}= C_i \exp \left(-\frac{(j-f D      x_{l,j}^{-1}-i)^2}{\sigma_s^2} \right)
 \label{eq:nweight}
\end{align}

Since $\g(\x_l)$ is differentiable, we use the first-order Taylor series expansion around $\x_l=\x_l^0$ to get a linear approximation, where $\x_l^0$ is the first estimate. Thus, 
\begin{align}
\g(\x_l) \approx ~ & \g(\x_l^0) + \g'(\x_l^0)(\x_l-\x_l^0) \\
= ~ & \g'(\x_l^0)\x_l + \g(\x_l^0)-\g'(\x_l^0)\x_l^0 \\
=~ & \H \x_l+ \d
\end{align}
where $\H = \g'(\x_l^0) = \left[ \frac{\partial g_i(\x_l^0)}{\partial x_{l,j}} \right]_{i,j} \in \mathbb{R}^{N \times N} $ is the \textit{Jacobian matrix} (first-order partial derivatives) of $\g(\x_l)$ at $\x_l=\x_l^0$, and $\d=\g(\x_l^0)-\g'(\x_l^0)\x_l^0$ is a constant vector.


\subsection{Likelihood Term}

We assume that the zero-mean additive noise $\n_l \in \mathbb{R}^N$ follows a jointly Gaussian distribution; \ie, the probability density function (pdf) of $\n_l$ is
\begin{align}
\rPr (\n_l) &= \exp \left( - \frac{\n_l^{\top} \P_l \n_l}{\sigma_n^2}\right)
\label{eq:noiseModel}
\end{align}
where $\P$ is a positive definite (PD)  precision matrix, and $\sigma_n^2$ is the noise variance. 
Given observation $\y_l$, the likelihood term $\rPr(\y_l | \x_l)$ is
\vspace{-0.1in}
\begin{align}
\rPr(\y_l | \x_l) = 
\int_{\cR_l} \rPr (\n_l) \; d \n_l
\label{eq:likelihood}
\end{align}
where the region $\cR_l$ over which the integration \eqref{eq:likelihood} is performed is defined as
\begin{align}
\cR_l = \left\{ \n_{l,i} ~|~ \y_{l,i} - \frac{Q}{2} < \x_{l,i} + \n_{l,i} < \y_{l,i} + \frac{Q}{2} \right\}
\end{align}
We assume that left and right noise $\n_l$ and $\n_r$ are independent.

\begin{figure}
\begin{center}
\includegraphics[width=2.8in]{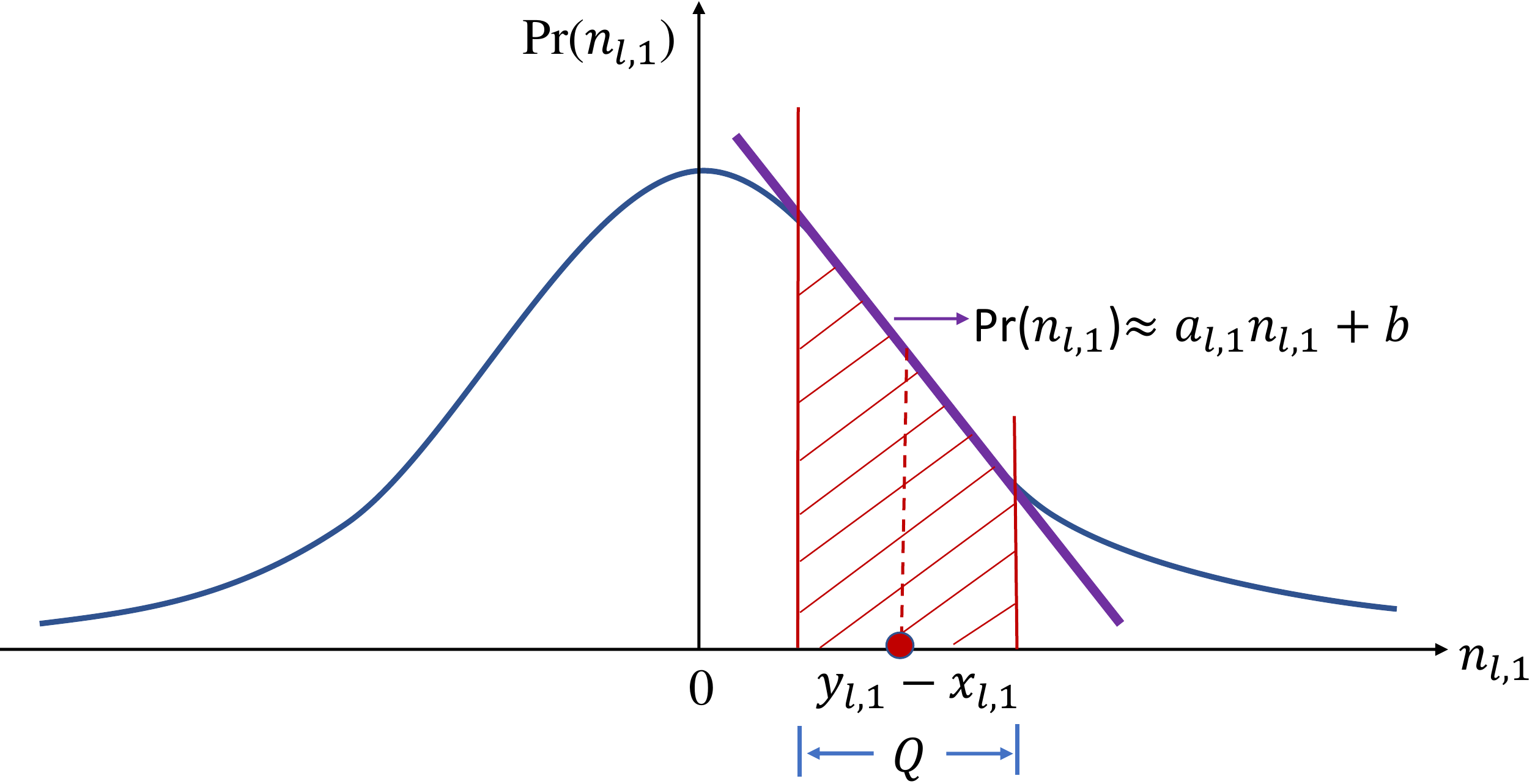}
\vspace{-0.1in}
\caption{\footnotesize Affine approximation of the Gaussian pdf.}
\label{fig:likelihood}
\end{center}
\vspace{-0.5cm}
\end{figure}

The integration in \eqref{eq:likelihood} over a jointly Gaussian pdf is non-trivial.
Instead, we first approximate $\rPr(\n_l)$ over the region $\cR_l$ as an affine function
\vspace{-0.1in}
\begin{align}
\rPr(\n_l) &\approx \a^{\top} \n_l + b
\end{align}
where contants $\a \in \mathbb{R}^N$ and $b \in \mathbb{R}$ can be computed via Taylor series expansion at $\y_l - \x_l$ given \eqref{eq:noiseModel}. 
See 1-D case in Fig.\,\ref{fig:likelihood} for an illustration.
For reasonably small $Q$, this is a good approximation. 
We now rewrite \eqref{eq:likelihood} as
\vspace{-0.05in}
\begin{align}
\rPr(\y_l | \x_l) &\approx
\int_{\cR_l} (\a^{\top} \n_l+ b) \; d \n_l   \\
&= Q^N \left( \a^{\top} (\y_l - \x_l) + b\right)
\label{eq:multiInt}
\end{align}
where \eqref{eq:multiInt} is proven in the Appendix.

\vspace{-0.2cm}
\subsection{Signal Prior}

As done in recent graph-based image processing work \cite{kalofolias2016learn, egilmez2017graph, bai2018graph}, we model the similarities among pixels in $\x_l$ using a graph Laplacian matrix $\L_l$, and thus prior $\rPr(\x_l)$ can be written as:
\begin{align}
\rPr(\x_l) = \exp \left( - \frac{\x_l^{\top} \L_l \x_l}{\sigma^2} \right)
\label{eq:prior}
\end{align}
We assume that the previous $K$ pixel rows in the left depth image have been enhanced, and assuming in addition that the next row $i$ follows a similar image structure, $\L_l$ can be learned from the previous $K$ rows. 
See Section\;\ref{sec:learn} for details.

\subsection{MAP Formulation}

We now formulate a MAP problem for $\widetilde{\X} = \{\widetilde{\x}_l, \widetilde{\x}_r\}$ as follows.
\begin{align}
 \max_{\x_l,\x_r} ~~& \rPr(\y_l,\y_r|\x_l,\x_r) \, \rPr(\x_l, \x_r) \\ 
=~& \rPr(\y_l,\y_r|\x_l,\g(\x_l)) \, \rPr(\x_l, \g(\x_l)) 
\label{eq:sub} \\
=~&  \rPr(\y_l|\x_l) \rPr(\y_r|\g(\x_l)) \, \rPr(\x_l) \rPr(\g(\x_l)) 
\label{eq:ind} \\
\approx ~& Q^{2N} (\a^{\top}(\y_l-\x_l)+b) (\a^{\top}(\y_r-\g(\x_l))+b) \nonumber\\
& \exp \left( - \frac{\x_l^{\top} \L_l \x_l}{\sigma^2} \right) \exp \left( - \frac{\g(\x_l)^{\top} \L_r \g(\x_l)}{\sigma^2} \right)
\label{eq:map}
\end{align}
where in \eqref{eq:sub} we substituted $\g(\x_l)$ for $\x_r$, and in \eqref{eq:ind} we split up the first term since left and right noise, $\n_l$ and $\n_r$, are independent.

To ease optimization, we minimize the negative log of \eqref{eq:map}:  

\vspace{-0.1in}
\begin{small}
\begin{align}
\min_{\x_l}  &~~ -\ln (\a^{\top}(\y_l-\x_l)+b)-\ln (\a^{\top}(\y_r-\g(\x_l))+b) \nonumber \\
& ~~ +\x_l^{\top} \L_l \x_l+ \g(\x_l)^{\top} \L_r \g(\x_l) \\
=& ~~ -\ln (\a^{\top}(\y_l-\x_l)+b)-\ln (\a^{\top}(\y_r-(\H \x_l+ \d))+b) \nonumber \\
& ~~ +\x_l^{\top} \L_l \x_l+ \x_l^{\top} \H^{\top} \L_r \H \x_l+2\d^{\top} \L_r \H \x_l+\d^{\top} \L_r \d
\label{eq:fmap}
\end{align}
\end{small}
\eqref{eq:fmap} is an unconstrained convex and differentiable objective; we can solve for its minimum efficiently using FGM.



\section{Feature Graph Learning}
\label{sec:learn}
\subsection{Learning Metric for Graph Construction}

When pixel row $i$ of the left view is optimized, we assume that the previous $K$ rows, $i-1, \ldots i-K$, have already been enhanced into $\widetilde{\x}_l^{i-1}, \ldots \widetilde{\x}_l^{i-K}$.
Using these $K$ enhanced rows, we compute graph Laplacian $\L_l$ to define prior $\rPr(\x)$ in \eqref{eq:prior}.
Because in a practice $K < N$, estimating $\L_l \in \mathbb{R}^{N \times N}$ reliably using only $K$ signal observations is a known difficult \textit{small data learning problem}.
In particular, established graph learning algorithms such as \textit{graphical LASSO} \cite{friedman08} and \textit{constrained $l_1$-norm minimization} (CLIME) \cite{cai11_CLIME} that compute a sparse precision matrix using as input an accurate empirical covariance matrix estimated from a large number of observations do not work in our small data learning scenario.

Instead, inspired by \cite{hu19} we construct an appropriate similarity graph via \textit{metric learning}.
We first assume that associated with each pixel (graph node) $i$ in $\x_l$ is a length-$F$ relevant \textit{feature vector} $\f_i \in \mathbb{R}^F$ (to be discussed). 
The \textit{feature distance} $d_{ij}$ between two nodes $i$ and $j$ is computed using a real, symmetric and PD \textit{metric} matrix $\M \in \mathbb{R}^{F \times F}$ as
\vspace{-0.05in}
\begin{align}
d_{ij} = (\f_i-\f_j)^{\top} \M (\f_i-\f_j)
\label{eq:featureDist}
\end{align}
Since $\M$ is PD, $d_{ij} > 0$ for $\f_i - \f_j \neq \0$. The edge weight $w_{ij}$ between nodes $i$ andd $j$ is then computed using a Gaussian kernel:
\vspace{-0.01in}
\begin{align}
w_{ij} = \exp \left( -d_{ij} \right)
\label{eq:edgeWeight}
\end{align}

To optimize $\M$, we minimize the \textit{graph Laplacian regularizer} (GLR) evaluated using $K$ previous pixel rows:
\begin{align} 
\min_{\M \succ 0} ~~ & \sum_{k=1}^K 
\left( \widetilde{\x}_l^k \right)^{\top} \L^k_l(\M) \widetilde{\x}_l^k 
\label{eq:metricLearn} \\
&= \sum_{k=1}^K \sum_{i,j} w^k_{ij} \left( \widetilde{x}_{l,i}^k - \widetilde{x}_{l,j}^k \right)^2
\end{align}
where edge weights $w_{ij}^k$ in Laplacian $\L^k_l(\M)$ is computed using features $\f_i^k$ and $\f_j^k$ of the $k$-th observation $\widetilde{\x}_l^k$ and equations \eqref{eq:featureDist} and \eqref{eq:edgeWeight}.
To optimize $\M$ in \eqref{eq:metricLearn}, \cite{hu19} proposed a fast optimization algorithm to optimize the diagonal and off-diagonal entries of $\M$ alternately. 
See \cite{hu19} for details.

\subsection{Feature Selection for Metric Learning}

To construct a feature vector $\f_i$ for each pixel $i$ in $\x_l$, we first compute the pixel's corresponding \textit{surface normal} $\n_i \in \mathbb{R}^3$ by projecting it to 3D space and computing it using its neighboring points via method \cite{avron2010}.
Then together with depth value $x_i$ and location $\l_i \in \mathbb{R}^2$ in the 2D grid, we construct $\f_i \in \mathbb{R}^6$.
Because $\M$ is symmetric, the number of matrix entries we need to estimate is only $21$.

\section{Experiments}
\label{sec:results}
\begin{table}
	\centering
	\caption{\footnotesize C2C and C2P results of competing methods at three noise levels.} \label{tab:level}
	\scriptsize
	\begin{tabular}{c|c|c|c|c|c|c}
		\Xhline{1.5pt}
		\textbf{$\sigma_n^2$} & methods & Adirondack & ArtL & Teddy & Recycle &Playtable\\
		\Xhline{1pt}
		\multirow{5}{*}{50} & 	\multirow{2}{*}{APSS} & 3.63 & 3.47 & 2.76 & 3.92 & 4.08  \\
		\cdashline{3-7}[0.8pt/2pt]
	      &  & 14.45 & 11.73 & 7.26 & 15.79  & 17.42 \\
	    \cline{2-7} 
	    & 	\multirow{2}{*}{RIMLS} & 3.47 & 3.35 & 2.67 & 3.72& 4.09 \\
	   \cdashline{3-7}[0.8pt/2pt]
	    &  & 13.26  & 11.21 & 7.09 & 15.11 & 17.06 \\
	    \cline{2-7} 
	    & 	\multirow{2}{*}{MRPCA} & 2.91 & 3.05 & 2.55 & 3.17 & 3.21  \\
	   \cdashline{3-7}[0.8pt/2pt]
	    &  & 8.86 & 8.73 & 6.21 & 10.21 & 9.17  \\
	    \cline{2-7}
	    &\multirow{2}{*}{Proposed} & \textbf{2.11}  & \textbf{2.26} & \textbf{1.56} & \textbf{2.45} &  \textbf{3.09}\\
	   \cdashline{3-7}[0.8pt/2pt]
	    &  & \textbf{4.88} & \textbf{6.34} & \textbf{2.79} & \textbf{7.00} & \textbf{8.78} \\
	    \Xhline{1pt}
		\multirow{5}{*}{70} & 	\multirow{2}{*}{APSS} & 4.12 & 3.80 & 3.09 & 4.34 & 4.46 \\
		\cdashline{3-7}[0.8pt/2pt]
	      &  & 18.56 & 13.88 & 8.96 & 20.07 & 18.91  \\
	    \cline{2-7} 
	    & 	\multirow{2}{*}{RIMLS} & 3.83 & 3.67 & 3.00 & 4.16 & 4.38 \\
	   \cdashline{3-7}[0.8pt/2pt]
	    &  & 17.26 & 13.41 & 8.73 & 19.42 & 18.92  \\
	    \cline{2-7} 
	    & 	\multirow{2}{*}{MRPCA} & 3.42 & 3.45 & 2.89 & 3.76 & 3.48 \\
	   \cdashline{3-7}[0.8pt/2pt]
	    &  & 12.57 & 11.39 & 7.98 & 14.80  &  11.00  \\
	    \cline{2-7}
	    &\multirow{2}{*}{Proposed} & \textbf{2.32} & \textbf{2.48} & \textbf{1.68} & \textbf{2.68} &  \textbf{3.26}\\
	   \cdashline{3-7}[0.8pt/2pt]
	    &  & \textbf{5.97} & \textbf{7.64} & \textbf{3.32} & \textbf{8.47} & \textbf{10.43} \\
	    \Xhline{1pt}
		\multirow{5}{*}{90} & 	\multirow{2}{*}{APSS} & 4.40 & 4.28 & 3.38 & 4.80 & 4.91  \\
		\cdashline{3-7}[0.8pt/2pt]
	      &  & 21.97 & 17.07& 11.08 & 25.11 & 26.18  \\
	    \cline{2-7} 
	    & 	\multirow{2}{*}{RIMLS} & 4.19 & 4.13& 3.30 & 4.59 & 4.83  \\
	   \cdashline{3-7}[0.8pt/2pt]
	    &  & 21.15  & 16.52 & 10.70& 24.16& 23.46 \\
	    \cline{2-7} 
	    & 	\multirow{2}{*}{MRPCA} & 3.78 &  3.91 & 3.20  & 4.20 & 3.95 \\
	   \cdashline{3-7}[0.8pt/2pt]
	    &  & 16.11 & 14.07 & 9.69 & 19.10  & 14.52  \\
	    \cline{2-7}
	    &\multirow{2}{*}{Proposed} & \textbf{2.47} & \textbf{2.70} & \textbf{1.84} & \textbf{2.92} &  \textbf{3.45} \\
	   \cdashline{3-7}[0.8pt/2pt]
	    &  & \textbf{6.95} &\textbf{9.15} & \textbf{4.08} & \textbf{10.45} & \textbf{13.22}  \\
       \Xhline{1.5pt}
    	\end{tabular}
    	\vspace{-0.4cm}
\end{table}

We conducted simulations with five depth image pairs provided in Middlebury datasets \cite{scharstein2014high}: \texttt{Adirondack},  \texttt{Recycle}, \texttt{Playtable},  \texttt{Teddy} and \texttt{ArtL}. 
By projecting left and right views to 3D space, the first three generate PCs with around 700000 points, \texttt{Teddy} with 337500 points and \texttt{ArtL} with 192238 points. 
Gaussian noise with zero mean and variance $\sigma_n^2$ of 50, 70 and 90 is added to both left and right views, which are then quantized into 256 distinct values. 
To compute the precision matrix $\P$ for noise in pixel row $i$, we use previous $K_n=30$ estimated noise terms to compute the covariance matrix, where $\widetilde{\n}^{i-k_n}=\y^{i-k_n}-\widetilde{\x}^{i-k_n}$ and $1 \leq k_n \leq K_n$. 
When learning metric for graph construction, we consider previous $K=10$ pixel rows. 
To reduce computation complexity, the same optimized $\M$ is used for the next $K$ pixel rows. 
Based on the feature vector in $\y^i$ of the current row $i$, we can finally compute the corresponding Laplacian $\L^i(\M)$.  

Our proposed 3D PC enhancement method is compared against three existing PC denoising algorithms: APSS \cite{guennebaud2007algebraic}, RIMLS \cite{oztireli2009feature} and the moving robust principle component analysis (MRPCA) algorithm \cite{mattei2017point}. APSS and RIMLS are implemented with MeshLab software, and the source code of MRPCA is provided by the authors. Two commonly used PC evaluation metrics, point-to-point (C2C) error and point-to-plane (C2P) error between ground truth and denoising point sets, are employed.

After projecting both noise-corrupted and quantized left and right views into a PC, one can employ three mentioned PC denoising algorithms. C2C and C2P results of different methods with three noise levels are shown in Table \ref{tab:level}. Overall, our method achieves by far the best performance in both metrics and all three noise levels, with C2C reduced by 0.68, 0.92, 1.13; and C2P reduced by 2.68, 4.38, 5.93 on average compared to the second best algorithm for $\sigma_n^2=$ 50, 70, 90, respectively. 

\begin{figure}[h]
\begin{center}
\includegraphics[width=3.4in]{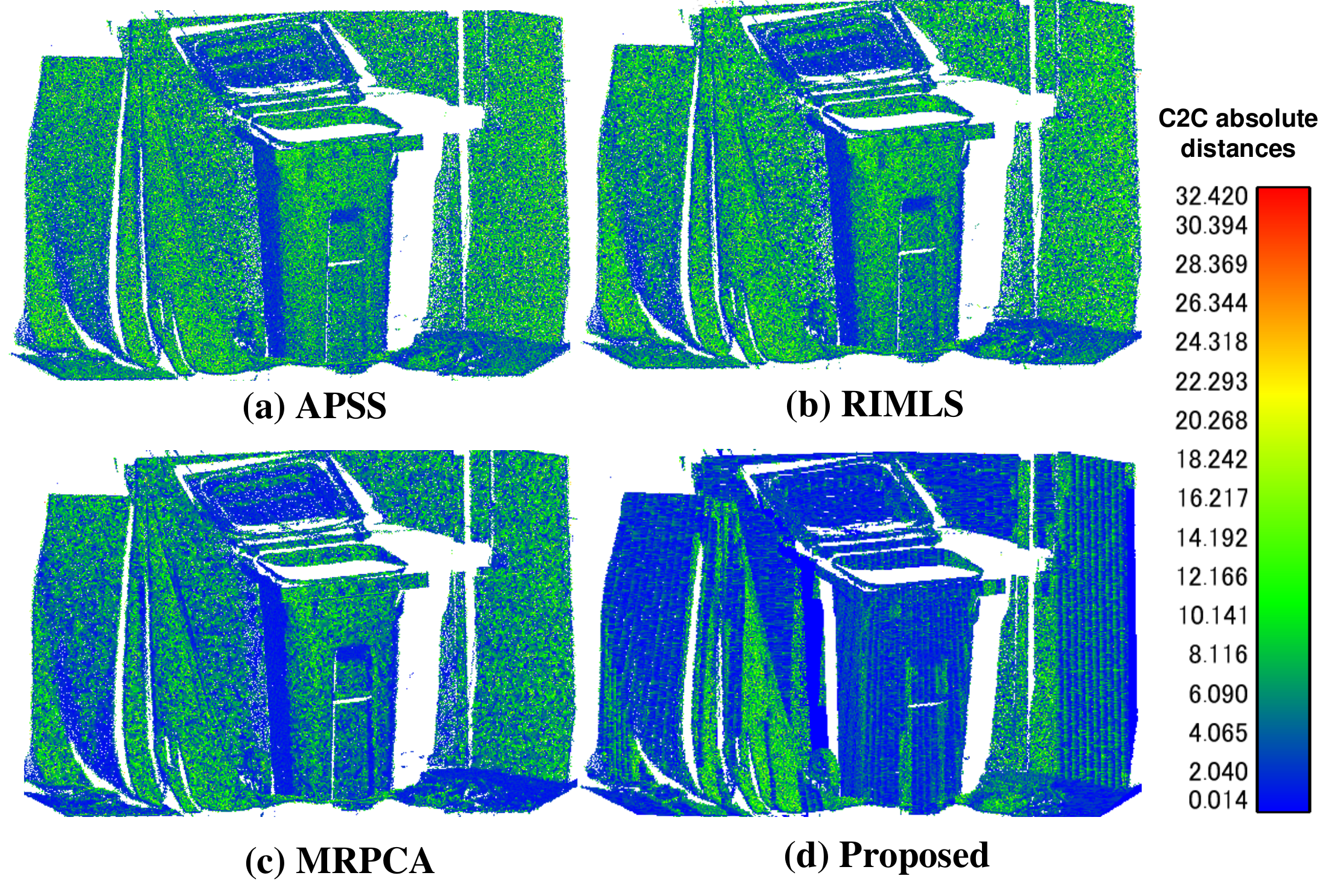}
\caption{\footnotesize Comparison of visual results for \texttt{Recycle} when $\sigma_n^2=$ 50. From blue to red, C2C absolute errors gradually become larger. More blue points are noticely included in the proposed method. }
\label{fig:visual}
\end{center}
\vspace{-0.6cm}
\end{figure}

Visual results for \texttt{Recycle} is shown in Fig.\,\ref{fig:visual}. 
For better visualization, we use CloudCompare software to show the C2C absolute distances between the ground truth points and their closest denoised points. 
We observe that our proposed method achieves smaller C2C errors (in blue) compared to the competitors. 

\section{Conclusion}
\label{sec:conclude}
Point clouds are typically synthesized from finite-precision depth measurements that are noise-corrupted.
In this paper, we improve the quality of a synthesized point cloud by jointly enhancing multiview depth images---the ``rawest" signal we can acquire from an off-the-shelf sensor---prior to modules in a typical point cloud synthesis pipeline that obscure acquisition noise. 
We formulate a graph-based MAP optimization that specifically targets an image formation model accounting for both additive noise and quantization. 
Simulation results show that our proposed scheme outperforms competing schemes that denoise point clouds after the synthesis pipeline.

\appendix
\section{Proof of Multiple Integral}

\vspace{-0.1in}
We prove \eqref{eq:multiInt} by induction. 
Consider first the base case ($N=1$) when $x_l, y_l, n_l \in \mathbb{R}$ and $\rPr(n_l) \approx a \, n_l + b$, where $a, b \in \mathbb{R}$. 
Integral in \eqref{eq:likelihood} in this case is a single integral, and one can easily check that
$\rPr(y_l | x_l) = Q \left(a (y_l - x_l) + b \right)$.
Consider next the inductive case and assume $\rPr(\y_l | \x_l) = Q^N \left(\a^{\top} (\y_l - \x_l) + b \right)$, when $\x_l, \y_l, \n_l \in \mathbb{R}^N$.
If the dimension of the signal is actually $N+1$, when integrating the first $N$ variables, the $N+1$-th term $a_{N+1} n_{N+1}$ is treated the same as constant $b$, thus,
\begin{small}
\begin{align}
\rPr(\y_l|\x_l) &= 
\int_{z_{N+1}-Q/2}^{z_{N+1}+Q/2} 
Q^N \left( \a'^{\top} \z'_l + b + a_{N+1} n_{N+1} \right) d n_{N+1}
\nonumber
\end{align}
\end{small}
\noindent
where $\a'$ and $\z'_l = \y_l - \x_l$ are vectors for only the first $N$ terms.
Since $\a'^{\top} \z'_l + b$ is constant, like the base case one can easily integrate this, resulting in
\begin{align}
\rPr(\y_l|\x_l) &= Q^N \left[ Q \left(a_{N+1} z_{N+1} + \a'^{\top} \z'_l + b \right) \right] \nonumber \\
&= Q^{N+1} \left( \a^{\top} \z_l + b \right) \nonumber 
\end{align}
where $\a$ and $\z_l$ are vectors for all $N+1$ terms. $\qed$


\bibliographystyle{IEEEbib}
\bibliography{ref2}

\end{document}